\documentclass[onecolumn,showkeys,preprintnumbers,aps,a4paper,amssymb,prd,superscriptaddress,nofootinbib]{revtex4-2}
\usepackage{comment}
\usepackage{graphicx}
\usepackage{epsf}
\usepackage{bm}
\usepackage{amsmath}
\usepackage{amsfonts}
\usepackage{amssymb}
\usepackage{epstopdf}
\usepackage{color}
\usepackage[dvipsnames]{xcolor}
\usepackage{verbatim}
\usepackage{multirow}
\usepackage{soul}
\usepackage{physics}
\usepackage{bm}

\usepackage[width=0.00cm, height=0.00cm, left=1.50cm, right=1.50cm, top=2.00cm, bottom=2.00cm]{geometry}
\usepackage{microtype}
\usepackage{lmodern}

\usepackage[colorlinks = true,
linkcolor = teal,
urlcolor  = teal,
citecolor = teal,
anchorcolor = blue]{hyperref}
\usepackage[capitalize]{cleveref}
\usepackage[normalem]{ulem}
\usepackage{enumitem}
\usepackage{booktabs}

\usepackage{lipsum}

\makeatletter\let\expandableinput\@@input\makeatother

\begin{document}
\begin{center}
		\vspace{0.4cm} {\large{\bf {Hu-Sawicki  $f(R)$ Gravity in a Non-Flat Universe: Constraints from DESI-DR2, BBN, and Type Ia Supernovae }} \\
		\vspace{0.4cm}
		\normalsize{Saurabh Verma$^1$, Manish Yadav$^2$,  Archana Dixit$^3$, Anirudh Pradhan$^4$, M. S. Barak$^5$ }\\
		\vspace{5mm}
		
        \normalsize{$^{1,2,5 }$ Department of Mathematics, Indira Gandhi University, Meerpur, Haryana 122502, India }\\ 
        \normalsize{$^{3}$ Department of Mathematics, Gurugram University, Gurugram, Haryana, India}\\

		\normalsize{$^{4 }$ Centre for Cosmology, Astrophysics and Space Science (CCASS), GLA University, Mathura-281406, Uttar Pradesh, India}\\ 

		\vspace{2mm}
		$^1$Email address: saurabh.math.rs@igu.ac.in\\
       $^2$Email address: manish.math.rs@igu.ac.in\\
            $^3$Email address: archana.ibs.maths@gmail.com\\
		$^4$Email address: pradhan.anirudh@gmail.com\\
        $^5$Email address: ms$_{-}$barak@igu.ac.in\\}
\end{center}

 
\maketitle
%
 {\bf Abstract}: 
 Late-time cosmic acceleration is conventionally ascribed to a cosmological constant, though $\Lambda$CDM continues to face several theoretical difficulties that keep alternative gravity models under active consideration. This work examines the Hu-Sawicki $f(R)$ model in a spatially non-flat background, with the modified-gravity parameter $b$ and the curvature density $\Omega_k$ both treated as free parameters, constrained using DESI-DR2 BAO, BBN, and four Type Ia supernova compilations -- PantheonPlus, PantheonPlus+SH0ES, 
Union3, and DESY5yr. Across all four combinations, $H_0$ and $\Omega_m$ stay close to their $\Lambda$CDM values, with a noticeable shift in $H_0$ appearing only for the SH0ES-calibrated 
dataset. The parameter $b$ departs from zero at better than $2\sigma$ in three of the four fits, most prominently for DESY5yr, whereas the SH0ES-calibrated combination alone prefers $b<0$. Curvature is where the results are most striking: with $b$ left free, $\Omega_k$ is 
consistent with flatness at $1\sigma$ throughout, in contrast to the $\Lambda$CDM fits on the same data, which mildly prefer a closed or open universe depending on the supernova sample used. A strong positive correlation among $b$, $H_0$, and $\Omega_k$ underlies this 
shift, suggesting that curvature signatures obtained under $\Lambda$CDM can be reabsorbed into the modified-gravity sector once this additional freedom is allowed. Statistical model 
comparison via AIC and BIC gives a mixed picture: AIC leans toward the Hu-Sawicki model in three of the four combinations, while BIC's heavier penalty on the extra parameter favors $\Lambda$CDM in most cases, and only DESY5yr is preferred under both criteria. These findings show that the Hu-Sawicki $f(R)$ scenario with unconstrained curvature is nonetheless a statistically feasible, if not obviously preferred, alternative to $\Lambda$CDM, and that curvature restrictions generated inside $\Lambda$CDM cannot be viewed as independent of the underlying gravity model.

\section{Introduction}

In addition to the inflationary epoch in the very early universe \cite{r1,r2}, cosmological observations show that the Universe is presently going through the phase of accelerated expansion. Despite a wealth of observational evidence, the physical origin of this late-time acceleration is one of the most fundamental open questions in modern cosmology. Within the standard $\Lambda$CDM model, this acceleration is interpreted as being due to the cosmological constant $(\Lambda)$, which serves as the dark energy, and fits many cosmological observations very well. However, there are a number of well-known theoretical and observational problems associated with the model, like the cosmological constant problem \cite{r3,r4} and the persisting Hubble tension \cite{r5}.
These problems have motivated the construction of some alternative cosmological models with dynamical dark energy \cite{r6}. Such models can be broadly grouped into two main categories. The first one maintains the framework of general relativity but adds an extra dark energy component with evolving properties \cite{r6,r7}. The second one is to explain the cosmic acceleration by changing the underlying theory of gravity itself, and the $f(R)$ gravity is one of the most studied and successful examples \cite{r8,r9,r10}. In this context, several investigations have explored extensions or modifications of general relativity as potential mechanisms for alleviating the $H_0$ tension, with recent summaries available in the literature (see Refs.\cite{rs1,rs2,rs3,rs4,rs5,rs6,rs7,rs8,rs9,rs10,rs11,rs12,rs13}). However, it has also been argued that a number of cosmic tensions are difficult to address simultaneously \cite{rs3}, and hence new models and independent observational tests are needed.\\

In $f(R)$ gravity, the Ricci scalar $R$ of the Einstein-Hilbert action is generalized to a function of the Ricci scalar $f(R)$. This is a simple but powerful extension of general relativity. In the past years, several viable $f(R)$ models have been proposed, such as the Starobinsky \cite{r11}, Hu-Sawicki \cite{r12}, Tsujikawa \cite{r13,r14}, and exponential \cite{r15,r16} models, all of them intended to explain the observed cosmic acceleration while being consistent with local gravity tests and cosmological observations \cite{r10,r17}. In general, a viable $f(R)$ model should satisfy several important conditions, including the positive effective gravitational coupling, the stability against cosmological perturbations, the convergence to the $\Lambda$CDM model in the high-curvature regime, the stability of the late-time de Sitter solution, the consistency with the equivalence principle, and so on. We focus in this work on the Hu-Sawicki $f(R)$ gravity model because it satisfies these viability conditions with a single extra free parameter, unlike the standard $\Lambda$CDM cosmological model. Its minimal extension makes it an attractive paradigm to consider possible deviations from General Relativity and face them with modern cosmological results.\\

The standard cosmic inflation is known to predict a flat geometry. This is because the curvature density parameter $\Omega_k$ decays exponentially during inflation but grows only as a power law afterward \cite{r20}. However, models of inflation leading to open \cite{r21,r22} or closed universes \cite{r23,r24} are also possible, although they often require some fine-tuning \cite{r20}. It has been suggested that spatial curvature may have been induced during the evolution of the universe once the growth of large-scale structure became non-linear. This conclusion was obtained based on the Silent Universe approximation \cite{r25}. The appearance of curvature might also be the secret to resolving the currently unresolved tension between CMB and distance-ladder estimates of the Hubble constant \cite{r26}.\\

In this perspective, the practice of setting $\Omega_{k,0}$, the present-day value of $\Omega_k$, to zero seems a little premature. There is also the possibility that more stringent constraints $\Omega_{k,0}$ could be important tests of eternal inflation models \cite{r27}. Furthermore, higher-order perturbations like second-order lensing corrections \cite{r27} and impacts from large-scale structure (such as local inhomogeneities) might skew our observations and cause the inferred value of $\Omega_{k,0}$ to deviate from the background value if they are not appropriately taken into account \cite{r27,r28}. Another aspect to bear in mind is the strong degeneracy which often exists between dark energy parameters and $\Omega_{k,0}$. Either setting the latter to zero or taking into account only particular classes of the former are common ways to get around the issue. The dark energy equation-of-state (EoS) parameter $w_{de}$, which often has a functional form, serves as an example. The result is that the study of spatial curvature is mostly done in a rather limited setting. It is a concern in particular that if the true value of $\Omega_{k,0}$ is not zero, then assuming a flat geometry will induce errors in $w_{de}$ which grow rapidly with redshift, even if the curvature is in fact only very small \cite{r29}.\\

Recent analyses of the Planck 2018 cosmic microwave background (CMB) observations within the $\Lambda$CDM framework have hinted that a mildly closed universe may be favored at the $99\%$ confidence level or so \cite{r30}. Motivated by this possibility, we study viable $f(R)$ gravity models without the assumption of spatial flatness and how such models are constrained by the current cosmological observations. In particular we compare the predictions of viable $f(R)$ gravity with the standard $\Lambda$CDM model by allowing the spatial curvature parameter, $\Omega_{K}$, to vary freely. However, the possibility of non-zero spatial curvature provides an opportunity to test whether modified gravity fits the observational data better than the standard cosmological model. To the best of our knowledge, a detailed analysis of the viable $f(R)$ gravity models in a non-flat universe has not been performed in the literature so far. In this work we study the exponential $f(R)$ gravity model as a viable scenario.\\

This is the framework for the rest of the paper. In Sec. \ref{s1} we present the theoretical framework of f(R) gravity in a cosmic context along with an introduction to the particular models considered in this study. Sec. \ref{s2} describes the techniques of parameter estimation and the observational data used in this study. In Sec. \ref{s3}, we present the main results of our study and the detailed explanation of the resulting observational constraints. Finally, a brief discussion of possible future research topics is contained in Sec. \ref{s4}, and the main findings from this study are summarized.

\section{$f(R)$ gravity and cosmology}
\label{s1}


The action of $f(R)$ gravity is given by \cite{r42}
\begin{align} \label{frAction}
S=\int d^4 x \frac{\sqrt{-g}}{2\kappa^2}f(R) +S_{M},
\end{align}
where $\kappa^2 = 8\pi G$, $G$ being Newton's constant, and $S_M$ is the matter action, covering both relativistic and non-relativistic components.

Varying \eqref{frAction} leads to the field equations of $f(R)$ gravity,
\begin{align} \label{eom}
FR_{\mu\nu}-\frac{1}{2}g_{\mu\nu}f -\nabla_{\mu}\nabla_{\nu}F+g_{\mu\nu}\square F=\kappa^2 T_{\mu\nu}^{(M)},
\end{align}
where $F\equiv df(R)/dR$, $\square \equiv g^{\mu\nu}\nabla_\mu\nabla_\nu$ is the d'Alembert operator, and $T_{\mu\nu}^{(M)}$ is the energy-momentum tensor for relativistic and non-relativistic matter. Eq.~\eqref{eom} can also be recast as
\begin{align}
    G_{\mu\nu} = \kappa^2 \bigg(T_{\mu\nu}^{(M)} + T_{\mu\nu}^{(de)}\bigg),
\end{align}
where $G_{\mu\nu}=R_{\mu\nu}-(1/2)g_{\mu\nu}R$ is the Einstein tensor and the dark-energy energy-momentum tensor is
\begin{align}
    T_{\mu\nu}^{(de)} = \frac{1}{\kappa^2}\bigg(G_{\mu\nu}-FR_{\mu\nu} + \frac{1}{2} g_{\mu\nu} f + \nabla_{\mu}\nabla_{\nu}F - g_{\mu\nu}\square F\bigg).
\end{align}

\textbf{Modified Friedmann equations.} We consider the spatially non-flat Friedmann-Lemaître-Robertson-Walker (FLRW) spacetime,
\begin{equation}\label{flrw}
ds^2=-dt^2+a^{2}(t)\bigg(\frac{dr^2}{1-Kr^2}+r^2 d \theta^2+r^2 \sin^2\theta\, d\phi^2\bigg),
\end{equation}
where $a(t)$ is the scale factor and $K=-1,0,1$ correspond to open, flat, and closed universes, respectively. Substituting \eqref{flrw} into \eqref{eom} gives the modified Friedmann equations,
\begin{align}
&3FH^2+\frac{3KF}{a^2}=\frac{1}{2}(FR-f)-3H\dot{F}+\kappa^2\rho_M,\label{feg1}\\
&\ddot{F}=H\dot{F}-2F\dot{H}+\frac{2KF}{a^2}-\kappa^2 (\rho_M+P_M), \label{feg2}
\end{align}
where $H=\dot a/a$ is the Hubble parameter, a dot denotes a derivative with respect to cosmic time $t$, and the Ricci scalar is
\begin{align}
R=12H^2+6\dot{H}+\frac{6K}{a^2}.
\end{align}

To examine the dark-energy sector and the role of spatial curvature separately, we rewrite Eqs.~\eqref{feg1} and \eqref{feg2} as
\begin{align}
\label{H2}
H^2&=\frac{\kappa^2}{3}(\rho_M+\rho_{de}+\rho_K),\\  
\label{Hdot}
\dot{H}&=-\frac{\kappa^2}{2}(\rho_M+\rho_{de}+\rho_K+P_M+P_{de}+P_K),
\end{align}
where $\rho_M=\rho_m+\rho_r$ is the total density of non-relativistic matter and radiation, while the dark-energy density and pressure read
\begin{align}
\label{eqrhode}
\rho_{de}&=\frac{3}{\kappa^2}\bigg(H^2(1-F)-\frac{1}{6}(f-FR)-H\dot{F}+\frac{K}{a^2}(1-F)\bigg),\\
P_{de}&=\frac{1}{\kappa^2}\bigg(\ddot{F}+2H\dot{F}+\frac{1}{2}(f-FR)-(1-F)\big(3H^2+2\dot{H}+\frac{K}{a^2}\big)\bigg).
\end{align}
Similarly, the effect of spatial curvature is captured by an effective density and pressure,
\begin{align}
\rho_K &=-\frac{3K}{\kappa^2 a^2}, \label{rhoK}\\
P_K &= \frac{K}{\kappa^2 a^2}.
\end{align}
Matter, radiation, dark energy, and curvature each satisfy their own continuity equation,
\begin{align}
    \frac{d \rho_{i}}{dt} + 3H(1+w_{i})P_{i}=0,
\end{align}
with equation-of-state parameters
\begin{align}
    w_{i} \equiv \frac{P_{i}}{\rho_{i}}\,, \qquad i = (m,r,de, K).
\end{align}

Expressing \eqref{H2} in terms of density parameters gives
\begin{align}
    1=\Omega_{m}+\Omega_{r}+\Omega_{de}+\Omega_{K}\,,
\end{align}
where
\begin{align}
\Omega_{i}=\frac{\kappa^2\rho_{i}}{3H^2}. 
\end{align}
From \eqref{rhoK}, $\Omega_{K}=-K/(aH)^2$, so that $\Omega_{K}>0$, $\Omega_K=0$, and $\Omega_K<0$ correspond to an open, flat, and closed universe, respectively.\\

The action of $f(R)$ gravity is given by
\begin{align} \label{frAction}
S=\int d^4 x \frac{\sqrt{-g}}{2\kappa^2}f(R) +S_{M},
\end{align}
where $\kappa^2 = 8\pi G$, $G$ being Newton's constant, and $S_M$ is the matter action, covering both relativistic and non-relativistic components.

Varying \eqref{frAction} leads to the field equations of $f(R)$ gravity,
\begin{align} \label{eom}
FR_{\mu\nu}-\frac{1}{2}g_{\mu\nu}f -\nabla_{\mu}\nabla_{\nu}F+g_{\mu\nu}\square F=\kappa^2 T_{\mu\nu}^{(M)},
\end{align}
where $F\equiv df(R)/dR$, $\square \equiv g^{\mu\nu}\nabla_\mu\nabla_\nu$ is the d'Alembert operator, and $T_{\mu\nu}^{(M)}$ is the energy-momentum tensor for relativistic and non-relativistic matter. Eq.~\eqref{eom} can also be recast as
\begin{align}
    G_{\mu\nu} = \kappa^2 \bigg(T_{\mu\nu}^{(M)} + T_{\mu\nu}^{(de)}\bigg),
\end{align}
where $G_{\mu\nu}=R_{\mu\nu}-(1/2)g_{\mu\nu}R$ is the Einstein tensor and the dark-energy energy-momentum tensor is
\begin{align}
    T_{\mu\nu}^{(de)} = \frac{1}{\kappa^2}\bigg(G_{\mu\nu}-FR_{\mu\nu} + \frac{1}{2} g_{\mu\nu} f + \nabla_{\mu}\nabla_{\nu}F - g_{\mu\nu}\square F\bigg).
\end{align}

\textbf{Modified Friedmann equations.} We consider the spatially non-flat Friedmann-Lemaître-Robertson-Walker (FLRW) spacetime,
\begin{equation}\label{flrw}
ds^2=-dt^2+a^{2}(t)\bigg(\frac{dr^2}{1-Kr^2}+r^2 d \theta^2+r^2 \sin^2\theta\, d\phi^2\bigg),
\end{equation}
where $a(t)$ is the scale factor and $K=-1,0,1$ correspond to open, flat, and closed universes, respectively. Substituting \eqref{flrw} into \eqref{eom} gives the modified Friedmann equations,
\begin{align}
&3FH^2+\frac{3KF}{a^2}=\frac{1}{2}(FR-f)-3H\dot{F}+\kappa^2\rho_M,\label{feg1}\\
&\ddot{F}=H\dot{F}-2F\dot{H}+\frac{2KF}{a^2}-\kappa^2 (\rho_M+P_M), \label{feg2}
\end{align}
where $H=\dot a/a$ is the Hubble parameter, a dot denotes a derivative with respect to cosmic time $t$, and the Ricci scalar is
\begin{align}
R=12H^2+6\dot{H}+\frac{6K}{a^2}.
\end{align}

To examine the dark-energy sector and the role of spatial curvature separately, we rewrite Eqs.~\eqref{feg1} and \eqref{feg2} as
\begin{align}
\label{H2}
H^2&=\frac{\kappa^2}{3}(\rho_M+\rho_{de}+\rho_K),\\  
\label{Hdot}
\dot{H}&=-\frac{\kappa^2}{2}(\rho_M+\rho_{de}+\rho_K+P_M+P_{de}+P_K),
\end{align}
where $\rho_M=\rho_m+\rho_r$ is the total density of non-relativistic matter and radiation, while the dark-energy density and pressure read
\begin{align}
\label{eqrhode}
\rho_{de}&=\frac{3}{\kappa^2}\bigg(H^2(1-F)-\frac{1}{6}(f-FR)-H\dot{F}+\frac{K}{a^2}(1-F)\bigg),\\
P_{de}&=\frac{1}{\kappa^2}\bigg(\ddot{F}+2H\dot{F}+\frac{1}{2}(f-FR)-(1-F)\big(3H^2+2\dot{H}+\frac{K}{a^2}\big)\bigg).
\end{align}
Similarly, the effect of spatial curvature is captured by an effective density and pressure,
\begin{align}
\rho_K &=-\frac{3K}{\kappa^2 a^2}, \label{rhoK}\\
P_K &= \frac{K}{\kappa^2 a^2}.
\end{align}
Matter, radiation, dark energy, and curvature each satisfy their own continuity equation,
\begin{align}
    \frac{d \rho_{i}}{dt} + 3H(1+w_{i})P_{i}=0,
\end{align}
with equation-of-state parameters
\begin{align}
    w_{i} \equiv \frac{P_{i}}{\rho_{i}}\,, \qquad i = (m,r,de, K).
\end{align}

Expressing \eqref{H2} in terms of density parameters gives
\begin{align}
    1=\Omega_{m}+\Omega_{r}+\Omega_{de}+\Omega_{K}\,,
\end{align}
where
\begin{align}
\Omega_{i}=\frac{\kappa^2\rho_{i}}{3H^2}. 
\end{align}
From \eqref{rhoK}, $\Omega_{K}=-K/(aH)^2$, so that $\Omega_{K}>0$, $\Omega_K=0$, and $\Omega_K<0$ correspond to an open, flat, and closed universe, respectively.\\

A physically acceptable $f(R)$ theory cannot be built from just any function. There are a few conditions a model has to satisfy: the effective gravitational constant should stay positive, and cosmological perturbations should remain stable. Written explicitly,
\begin{eqnarray}
\label{13}
f_{,R} > 0 \quad \text{and} \quad f_{,RR} > 0 \quad 
\text{for} \quad R \geq R_0 > 0,
\end{eqnarray}
with $R_0$ the Ricci scalar at the present epoch. Ghost instabilities are avoided when $f_{,R}>0$, and the model is protected against a tachyonic instability when $f_{,RR}>0$~\cite{r43}. This is not the whole story, though, since a viable model must also agree with observations. At high curvature, for instance, it needs to converge to the $\Lambda$CDM limit,
\begin{eqnarray}
\label{14}
f(R) \rightarrow R - 2\Lambda, \quad \text{for} \quad 
R \geq R_0,
\end{eqnarray}
so that the matter-dominated epoch comes out right and the model stays compatible with the equivalence principle and Solar System tests. There is one more requirement: a stable de Sitter point at late times, for which
\begin{eqnarray}
\label{15}
0 < \left(\frac{R\, f_{,RR}}{f_{,R}}\right)_{r} < 1 
\quad \text{at} \quad r = -\frac{R\, f_{,R}}{f} = -2.
\end{eqnarray}
Once all these are imposed, viable $f(R)$ models with at most two free parameters take the general form
\begin{eqnarray}
\label{16}
f(R) = R - 2\, y(R, b)\, \Lambda,
\label{e16}
\end{eqnarray}

where $y(R,b)$ describes the departure of the model from general relativity, and $b$, a dimensionless parameter, fixes how large this departure is. The Hu-Sawicki $f(R)$ model~\cite{r44} is the specific case we study in this work.\\

\textbf{The Hu–Sawicki $f(R)$ model:}

Among the various functional forms proposed within $f(R)$ gravity, the model put forward by Hu and Sawicki~\cite{r44} has received considerable attention owing to its ability to pass solar system tests, which many earlier $f(R)$ proposals failed to satisfy. Instead of introducing a cosmological constant by hand, the late-time cosmic acceleration in this model arises from a nonlinear modification of the gravitational Lagrangian. The form is chosen so that it behaves like a cosmological constant at high curvature while recovering standard general relativity at low curvature, which keeps the model viable across both early and late cosmic epochs. The model is defined as
\begin{eqnarray}
\label{17}
f(R) = R - \frac{c_1\, R_{\mathrm{HS}} 
\left(R/R_{\mathrm{HS}}\right)^p}{c_2 
\left(R/R_{\mathrm{HS}}\right)^p + 1},
\label{HSfR}
\end{eqnarray}
where $c_1$, $c_2$, $R_{\mathrm{HS}}$, and $p>0$ are free parameters. Writing Eq.~\eqref{HSfR} in the general form of Eq.~\eqref{e16}, the distortion function $y(R,b)$ becomes
\begin{equation}
\label{18}
y(R, b) = 1 - \frac{1}{1 + 
\left(\dfrac{R}{\Lambda b}\right)^p},
\end{equation}
where $c_1 R_{\mathrm{HS}}/2c_2 = \Lambda$ and $2c_2^{1-1/p}/c_1 = b$ give the mapping between the two sets of parameters. Throughout this work, $p=1$ is held fixed this choice is standard in the literature, as $p$ is degenerate with $\Lambda$ and $b$, and the data alone cannot fix its value separately. Taking $b \rightarrow 0$, which corresponds to $c_1 \rightarrow \infty$ and $R_{\mathrm{HS}} \rightarrow 0$ while $c_1 R_{\mathrm{HS}}$ is held at $2\Lambda c_2$, one recovers $\Lambda$CDM exactly, with $f(R) \rightarrow R - 2\Lambda$.

\section{Datasets and Methodology}
\label{s2}

\subsection{\textbf{DESI BAO DR-2}}
We use baryon acoustic oscillation (BAO) measurements from the second data release of the Dark Energy Spectroscopic Instrument (DESI-DR2), combining Lyman-$\alpha$ forest tracers~\cite{r45} with galaxy and quasar observations~\cite{r46}. The dataset spans nine redshift bins over $0.295 \leq z \leq 2.330$ and is summarized in Table IV of Ref.~\cite{r46}. For tracers where only the isotropic BAO signal could be extracted, we quote the volume-averaged distance $D_V/r_d$; where the anisotropic decomposition is possible, the transverse comoving distance $D_M/r_d$ and the Hubble distance $D_H/r_d$ are used instead, with $r_d$ denoting the comoving sound horizon at the drag epoch in all cases. Because $D_V/r_d$, $D_M/r_d$, and $D_H/r_d$ are not statistically independent, we additionally include the correlation coefficients $r_{V,M/H}$ (between $D_V/r_d$ and $D_M/D_H$) and $r_{M,H}$ (between $D_M/r_d$ and $D_H/r_d$), so that these correlations are properly accounted for in the analysis. We refer to this dataset as \emph{DESI-DR2} throughout.

\subsection{\textbf{Big Bang Nucleosynthesis}}
 We also impose Big Bang Nucleosynthesis (BBN) constraints, based on the observed primordial abundances of light elements. In particular, we use the measured deuterium abundance, $y_{DP} = 10^{5}\, n_{D}/n_{H}$~\cite{r47}, together with the helium mass fraction $Y_{P}$~\cite{r48}. Since the BBN likelihood depends on the effective number of relativistic neutrino species $N_{\rm eff}$, and mainly constrains the physical baryon density $\omega_b \equiv \Omega_b h^2$, we fix $N_{\rm eff} = 3.046$ to its standard value throughout. Theoretical predictions for the primordial abundances are obtained using the PArthENoPE~2.0 code~\cite{r49}. We refer to this dataset as \emph{BBN} throughout.

 \subsection{Type Ia Supernovae (SN Ia)}

In this analysis, we make use of several recent Type Ia supernova (SN Ia) compilations, summarized below:

\begin{enumerate}
    \item \textbf{PantheonPlus and PantheonPlus+SH0ES:}
    
The PantheonPlus compilation provides distance-modulus measurements for 1701 light curves from 1550 distinct SN Ia, spanning $0.01 \leq z \leq 2.26$~\cite{r50}. We denote this dataset \emph{PP} in what follows. To allow a direct determination of the SN Ia absolute magnitude, we also consider a calibrated version of this sample that incorporates the latest SH0ES Cepheid host-galaxy distance anchors~\cite{r51}. Calibrating supernova luminosities directly through Cepheid distances, rather than simply imposing an $H_0$ prior from SH0ES, offers a more robust basis for cosmological inference. This calibrated dataset is referred to as \emph{PPS}.

\begin{figure}[hbt!]
    \centering
    \includegraphics[width=0.9\linewidth]{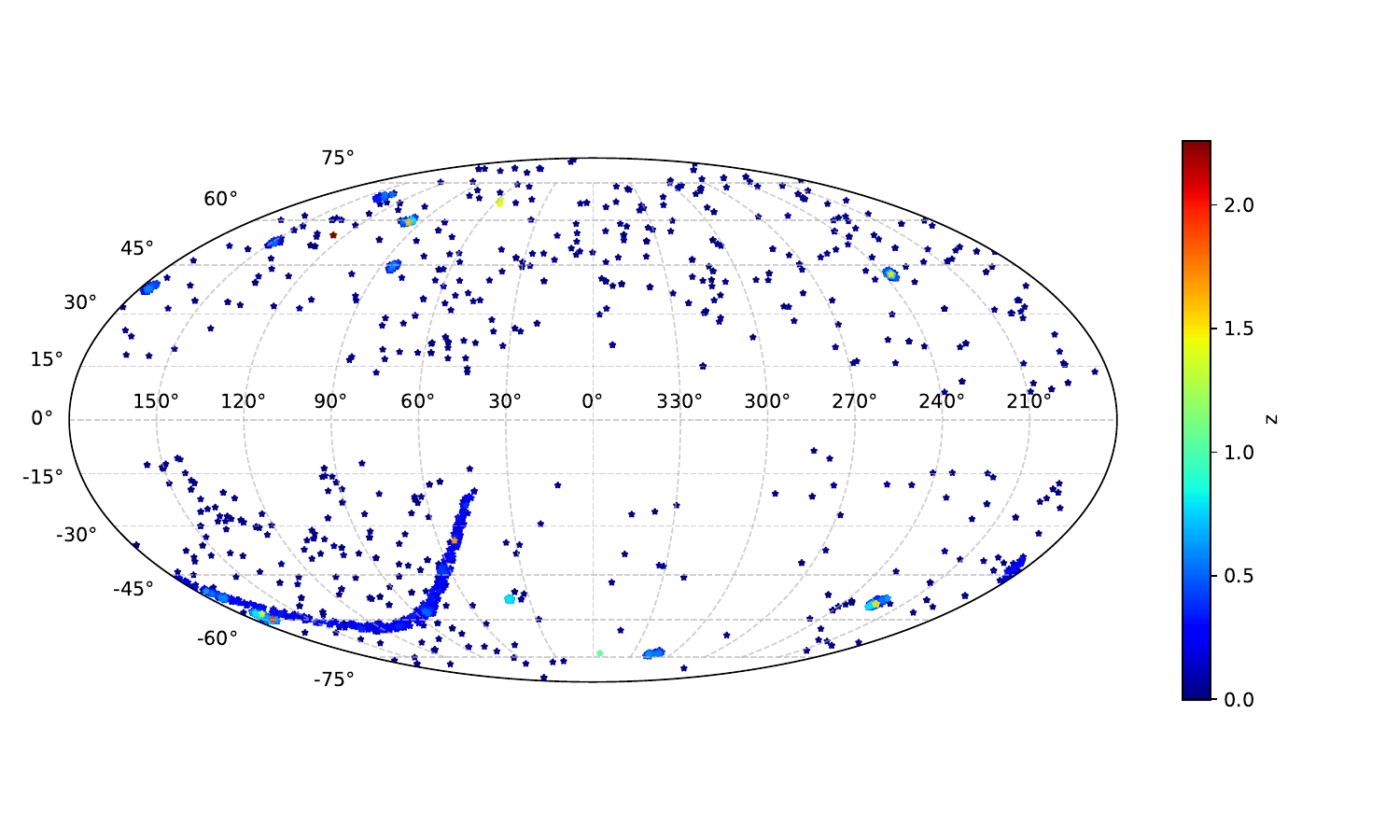}
    \caption{Sky distribution, in galactic coordinates, of the 1701 Type Ia supernovae in the PantheonPlus+SH0ES (PPS) compilation, color-coded by redshift $z$. }
    \label{f1}
\end{figure}

\item \textbf{Union 3.0:}

The Union 3.0 compilation consists of 2087 SN Ia over the range $0.001 < z < 2.260$~\cite{r52}, of which 1363 overlap with the PantheonPlus sample. A key strength of this compilation lies in its treatment of observational uncertainties through a Bayesian hierarchical modeling approach. We refer to this dataset as \emph{Union3}.

\begin{figure}[hbt!]
    \centering
    \includegraphics[width=0.9\linewidth]{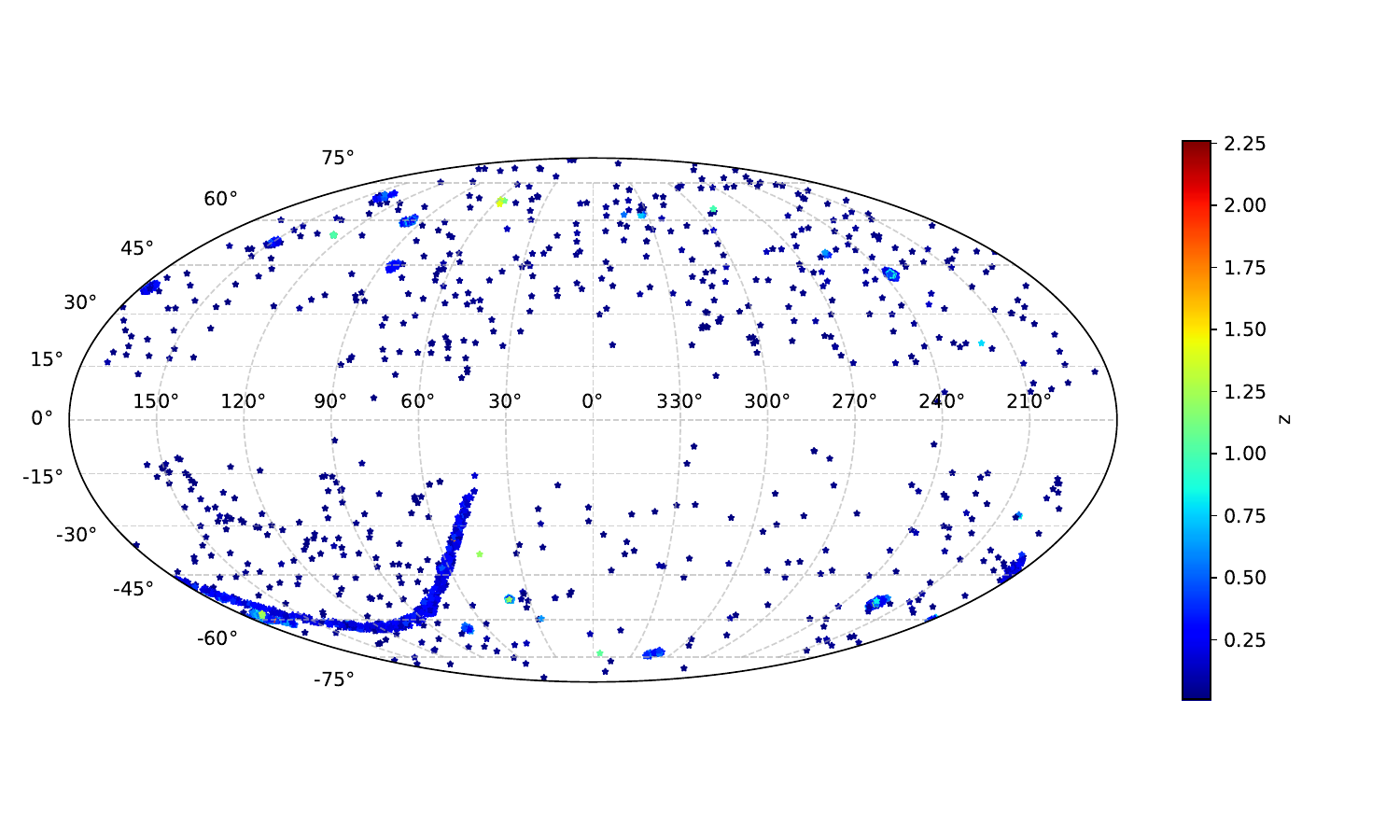}
    \caption{Sky distribution, in galactic coordinates, of the Union3 compilation, color-coded by redshift $z$.}
    \label{f2}
\end{figure}

 \item \textbf{DESY5:}
    
The Dark Energy Survey (DES) Year 5 release provides a newly assembled, homogeneously selected sample of 1635 photometrically classified SN Ia spanning $0.1 < z < 1.3$~\cite{r53}. An additional 194 low-redshift SN Ia in the range $0.025 < z < 0.1$, drawn from the overlap with PantheonPlus, are appended to this compilation. We label this dataset \emph{DESY5}.

\begin{figure}[hbt!]
    \centering
    \includegraphics[width=0.9\linewidth]{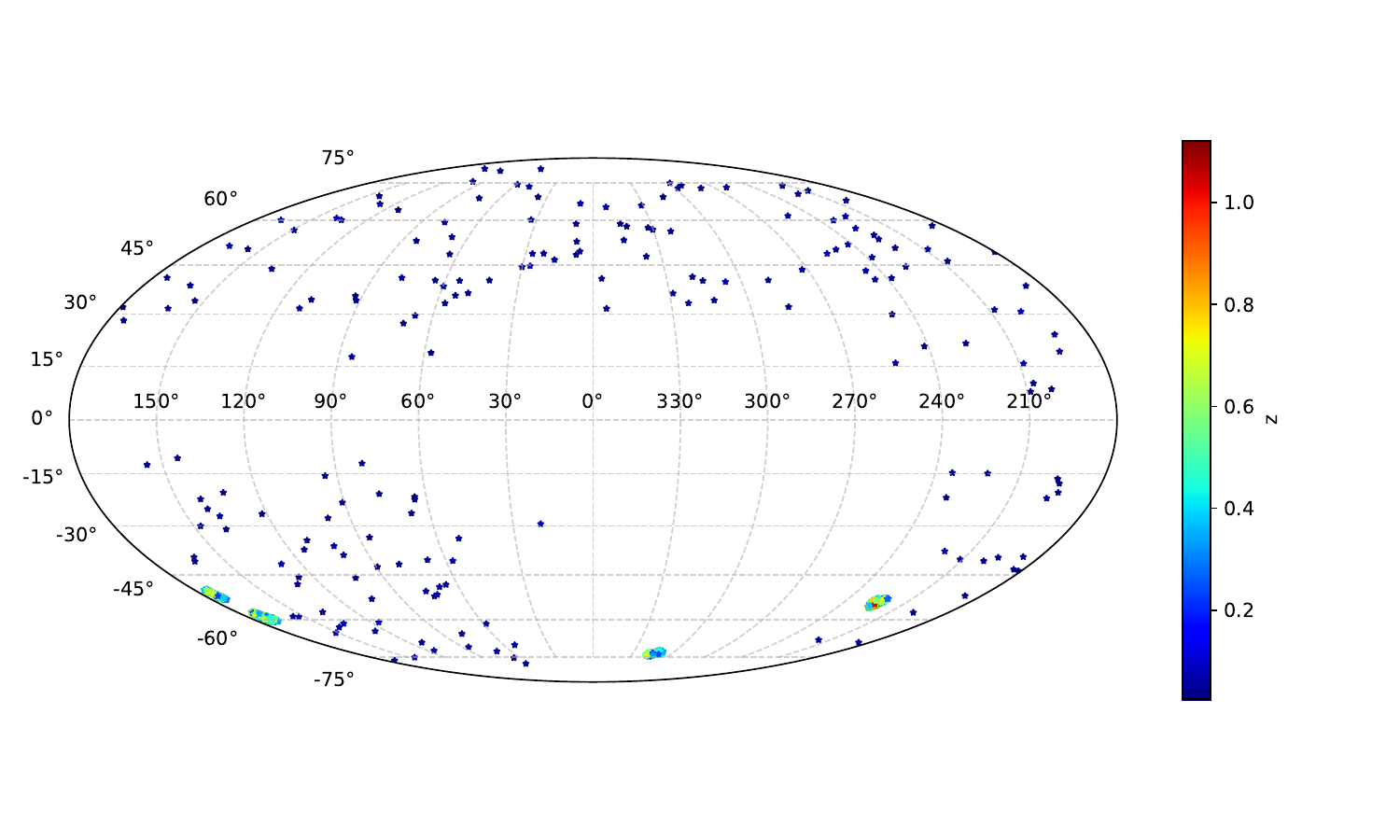}
    \caption{Sky distribution, in galactic coordinates, of the DESY5yr compilation, color-coded by redshift $z$.}
    \label{f3}
\end{figure}
\end{enumerate}

All cosmological observables in this work are computed using the Boltzmann solver \texttt{CLASS}~\cite{r54, r55}. To obtain observational constraints on the models considered here, we interface \texttt{CLASS} with the Markov Chain Monte Carlo sampler \texttt{MontePython}~\cite{r56}, suitably modified to accommodate the $f(R)$ gravity scenarios studied in this work.

\section{Result and Discussion}
\label{s3}

Table \ref{t1} summarizes the constraints on the Hu-Sawicki $f(R)$ model and its $\Lambda$CDM counterpart obtained from four data combinations: DESI-DR2+BBN+PP, DESI-DR2+BBN+PPS, 
DESI-DR2+BBN+Union3, and DESI-DR2+BBN+DESY5yr. Across all four combinations, the Hubble constant returned by the Hu-Sawicki model is consistent with the corresponding $\Lambda$CDM value within $1\sigma$, except for the PPS combination, where the SH0ES calibration pulls $H_0$ toward $71.9 \pm 2.0$ km/s/Mpc for the modified-gravity model against $72.20 \pm 0.8$ km/s/Mpc for $\Lambda$CDM. This behavior reflects the known sensitivity of $H_0$ to the SN Ia absolute-magnitude calibration rather than an intrinsic feature of the $f(R)$ sector, since the uncalibrated PP, Union3, and DESY5yr 
combinations all cluster around $H_0 \sim 67$--$68$ km/s/Mpc, close to the Planck-inferred value.\\

\begin{figure}[hbt!]
    \centering
    \includegraphics[width=1\linewidth]{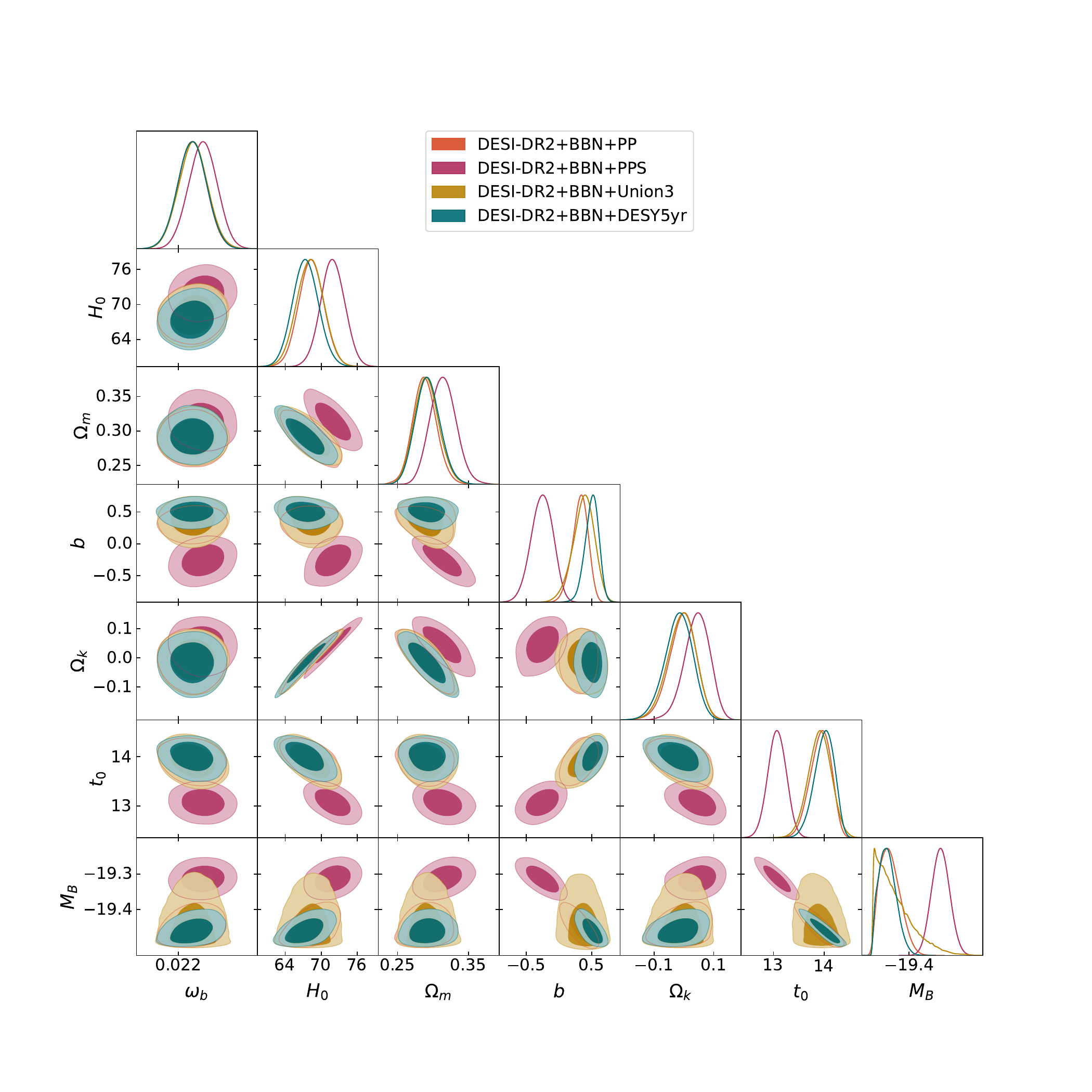}
    \caption{Two-dimensional marginalized posterior contours at 68\% and 95\% CL for the cosmological parameters  obtained from the Hu-Sawicki $f(R)$ gravity model using four dataset combinations: 
    DESI-DR2+BBN+PP, DESI-DR2+BBN+PPS, DESI-DR2+BBN+Union3, and DESI-DR2+BBN+DESY5. The diagonal panels show the one-dimensional marginalized posterior distributions for each parameter.}
    \label{f4}
\end{figure}

For the Hu-Sawicki model, the matter density parameter $\Omega_m$ stays fairly 
stable across the four combinations, falling between $0.288$ and $0.314$, and 
comes out marginally lower than the corresponding $\Lambda$CDM value in every 
single case. This small downward shift can be traced to the degeneracy between 
$\Omega_m$ and the distortion parameter $b$, which shows up clearly in the 
contours of Fig.~\ref{f4}: when $b$ moves further away from zero, $\Omega_m$ adjusts 
downward so that the late-time expansion history required by the BAO and SN 
distance measurements are left essentially unchanged. Turning to the Hu-Sawicki parameter $b$ itself, which measures how far the model 
sits from $\Lambda$CDM ($b \to 0$ being the limit where the concordance model is 
recovered), the constraints obtained are $b = 0.32^{+0.13}_{-0.11}$ for PP, 
$b = -0.26^{+0.18}_{-0.16}$ for PPS, $b = 0.37^{+0.18}_{-0.14}$ for Union3, and 
$b = 0.501^{+0.11}_{-0.092}$ for DESY5yr. In three out of four cases the data push 
$b$ toward positive values that differ from zero by more than $2\sigma$, and this 
departure is most pronounced for DESY5yr. PPS breaks this pattern, coming out with 
a negative $b$ instead — a result that can be linked back to the way the SH0ES 
calibration shifts $H_0$ upward and, with it, changes the late-time expansion rate 
the fit is trying to match. Because the sign and size of $b$ shift depending on 
which SN Ia sample is used, it is fair to say the present data do not yet pin down 
the departure from $\Lambda$CDM in any unique way, and the choice of SN calibration 
clearly has a real effect on how strong that departure looks. When the spatial curvature $\Omega_k$ is also left free, none of the four combinations 
show any meaningful deviation from flatness for the Hu-Sawicki model — every case is 
consistent with $\Omega_k = 0$ within $1\sigma$, with central values ranging from 
$-0.018$ to $+0.044$ and error bars around $0.04$--$0.05$. This is notably different 
from what happens under $\Lambda$CDM, where PPS instead points toward a closed universe 
($\Omega_k = -0.063 \pm 0.020$, a deviation from flatness just past $3\sigma$) and 
DESY5yr leans mildly toward an open one ($\Omega_k = 0.070^{+0.032}_{-0.026}$).\\

\begin{table}[hbt!]

\caption{
The ``Hu-Sawicki $f(R)$, and $\Lambda$CDM" models acquired from the DESI-DR2+BBN+PP, DESI-DR2+BBN+PPS, DESI-DR2+BBN+Union3, and DESI-DR2+BBN+DESY5yr datasets have constraints at 68$\%$ and 95$\%$ CL on a few chosen parameters.}
\vspace{0.5cm}
\label{t1}

\centering
\resizebox{\textwidth}{!}{
\begin{tabular}{ c | c | c | c | c }
\hline
\textbf{Data} & \textbf{DESI-DR2+BBN+PP} & \textbf{DESI-DR2+BBN+PPS} & \textbf{DESI-DR2+BBN+Union3} & \textbf{DESI-DR2+BBN+DESY5yr} \\
\hline

\textbf{Model} & \textbf{Hu-Sawicki $f(R)$} & \textbf{Hu-Sawicki $f(R)$} & \textbf{Hu-Sawicki $f(R)$} & \textbf{Hu-Sawicki $f(R)$} \\
& \textcolor{purple}{\textbf{$\Lambda$CDM}} & \textcolor{purple}{\textbf{$\Lambda$CDM}} & \textcolor{purple}{\textbf{$\Lambda$CDM}} & \textcolor{purple}{\textbf{$\Lambda$CDM}} \\
\hline

$H_0\,[{\rm km}/{\rm s}/{\rm Mpc}]$
& $68.4\pm 2.1$
& $71.9\pm 2.0$
& $68.2\pm 2.2$
& $67.4\pm 2.1$ \\
& \textcolor{purple}{$67.11^{+0.96}_{-1.2}$}
& \textcolor{purple}{$72.20\pm 0.8$}
& \textcolor{purple}{$67.7\pm 1.4$}
& \textcolor{purple}{$66.41^{+0.85}_{-1.3}$} \\
\hline

$\Omega_{\rm m}$
& $0.288\pm 0.017$
& $0.314\pm 0.018$
& $0.291^{+0.016}_{-0.018}$
& $0.293^{+0.016}_{-0.018}$
 \\
& \textcolor{purple}{$0.2940\pm 0.0097$}
& \textcolor{purple}{$0.3224\pm 0.0090$}
& \textcolor{purple}{$0.298\pm 0.011$}
& \textcolor{purple}{$0.2984^{+0.0090}_{-0.010}$} \\
\hline

$M_B$
& $-19.452^{+0.021}_{-0.034}$ 
& $-19.313\pm 0.025$
& $-19.438^{+0.016}_{-0.061}$
& $-19.458^{+0.018}_{-0.028}$
 \\
& \textcolor{purple}{$-19.447^{+0.029}_{-0.037}$}
& \textcolor{purple}{$-19.300\pm 0.024$}
& \textcolor{purple}{$-19.427^{+0.019}_{-0.072}$}
& \textcolor{purple}{$-19.453^{+0.024}_{-0.037}$} \\
\hline

$10^{2}\omega_b$
& $2.235\pm 0.00035$
& $2.261\pm 0.000355$
& $2.237\pm 0.0003$
& $2.235\pm 0.0003$
 \\
& \textcolor{purple}{$2.234\pm 0.00036$}
& \textcolor{purple}{$2.254\pm 0.00035$}
& \textcolor{purple}{$2.237\pm 0.00035$}
& \textcolor{purple}{$2.239\pm 0.00035$} \\
\hline

$\omega_{cdm}$
& $0.1121^{+0.0049}_{-0.0067}$
& $0.1397^{+0.0064}_{-0.0072}$
& $0.1127^{+0.0058}_{-0.0069}$
& $0.1104^{+0.0045}_{-0.0062}$
 \\
& \textcolor{purple}{$0.1102^{+0.0068}_{-0.0085}$}
& \textcolor{purple}{$0.1456\pm 0.0073$} 
& \textcolor{purple}{$0.1146^{+0.0093}_{-0.010}$}
& \textcolor{purple}{$0.1093^{+0.0062}_{-0.0088}$}\\
\hline

$b$
& $0.32^{+0.13}_{-0.11}$
& $-0.26^{+0.18}_{-0.16}$
& $0.37^{+0.18}_{-0.14}$
& $0.501^{+0.11}_{-0.092}$
 \\
& \textcolor{purple}{$-$}
& \textcolor{purple}{$-$}
& \textcolor{purple}{$-$}
& \textcolor{purple}{$-$} \\
\hline

$\Omega_k$
& $-0.005^{+0.048}_{-0.041}$
& $0.044^{+0.046}_{-0.038}$
& $-0.008^{+0.050}_{-0.042} $
& $-0.018^{+0.050}_{-0.042}$
 \\
& \textcolor{purple}{$0.046^{+0.029}_{-0.026}$}
& \textcolor{purple}{$-0.063\pm 0.020$}
& \textcolor{purple}{$0.032\pm 0.035 $}
& \textcolor{purple}{$0.070^{+0.032}_{-0.026}$} \\
\hline

$t_{0}$
& $13.93^{+0.23}_{-0.18}$
& $13.07\pm 0.18$
& $13.91\pm 0.23$
& $14.00^{+0.22}_{-0.17}$
 \\
& \textcolor{purple}{$13.92^{+0.23}_{-0.20}$}
& \textcolor{purple}{$13.05\pm 0.17$}
& \textcolor{purple}{$13.81\pm 0.264$}
& \textcolor{purple}{$13.91^{+0.24}_{-0.18}$} \\
\hline

$\chi^{2}_{min}$
& $1418.34$
& $1314.98$
& $33.12$
& $1668.5$ \\
& \textcolor{purple}{$1421.48$}
& \textcolor{purple}{$1312.08$}
& \textcolor{purple}{$37.28$}
& \textcolor{purple}{$1678.28$} \\
\hline
\end{tabular}}
\end{table}

Fig.~\ref{f4} shows the full triangular posterior for the Hu-Sawicki $f(R)$ model, combining the one-dimensional marginalized distributions and two-dimensional 68\% and 95\% confidence contours for $\omega_b$, $H_0$, $\Omega_m$, $b$, $\Omega_k$, $t_0$, and $M_B$, across the four data combinations. As a direct result of the SH0ES calibration anchoring the SN Ia absolute magnitude, the PPS contours are clearly moved toward greater $H_0$ and $\Omega_m$ in comparison to the other three combinations. In contrast, the PP, Union3, and DESY5yr contours cluster around $H_0 \simeq 67$--$68$ km/s/Mpc and overlap significantly in the $H_0$--$\Omega_m$ plane.  A clear negative correlation between $H_0$ and $\Omega_m$ is present in every combination, as expected from the BAO distance constraints, while the $b$--$\Omega_k$ panel already hints at the strong positive correlation between the modified-gravity parameter and spatial curvature that we discuss in more detail below.\\

\begin{figure}[hbt!]
    \centering
    \includegraphics[width=0.6\linewidth]{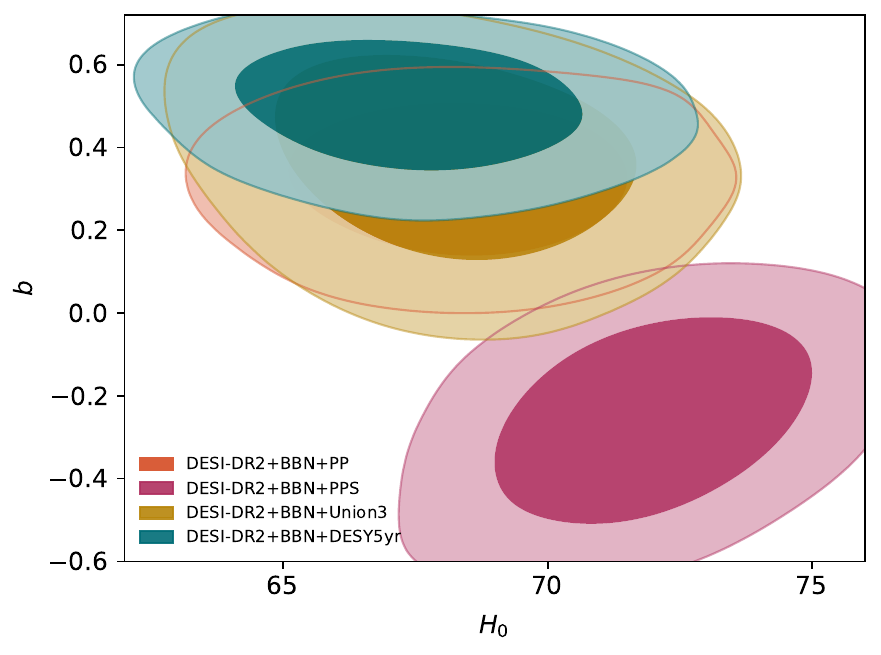}
    \caption{Two-dimensional marginalized posterior contours in the $(b-H_0)$ plane at 68\% and 95\% CL for the Hu-Sawicki $f(R)$ gravity model using four dataset combinations: 
    DESI-DR2+BBN+PP, DESI-DR2+BBN+PPS, DESI-DR2+BBN+Union3, and DESI-DR2+BBN+DESY5.}
    \label{f5}
\end{figure}

Fig.~\ref{f5} shows the two-dimensional 68\% and 95\% confidence-level contours in the 
$H_0$--$b$ plane for the Hu-Sawicki $f(R)$ model, across the four DESI-DR2+BBN+SN 
combinations. The PPS contour is clearly separated from the other three, 
occupying a distinct region at higher $H_0 \sim 68$--$75$ km/s/Mpc and negative $b$, 
centered near $b \sim -0.25$, consistent with the SH0ES-calibrated $H_0$ value reported 
in Table~\ref{t1}. The PP, Union3, and DESY5yr contours instead cluster together at lower $H_0 \sim 63$--$70$ km/s/Mpc and positive $b$, with DESY5yr occupying the highest region of $b$, extending up to $b \sim 0.6$--$0.7$, and showing substantial overlap 
with the Union3 contour. Within this cluster, no strong internal correlation between $H_0$ 
and $b$ is visible, since the contours are oriented nearly horizontally rather than 
diagonally. The clear separation between the PPS contour and the other three combinations 
indicates that the preferred sign of the departure from $\Lambda$CDM is driven mainly by the SN Ia calibration rather than by the BAO or BBN data, which is common to all four combinations.\\

\begin{figure}[hbt!]
    \centering
    \includegraphics[width=0.6\linewidth]{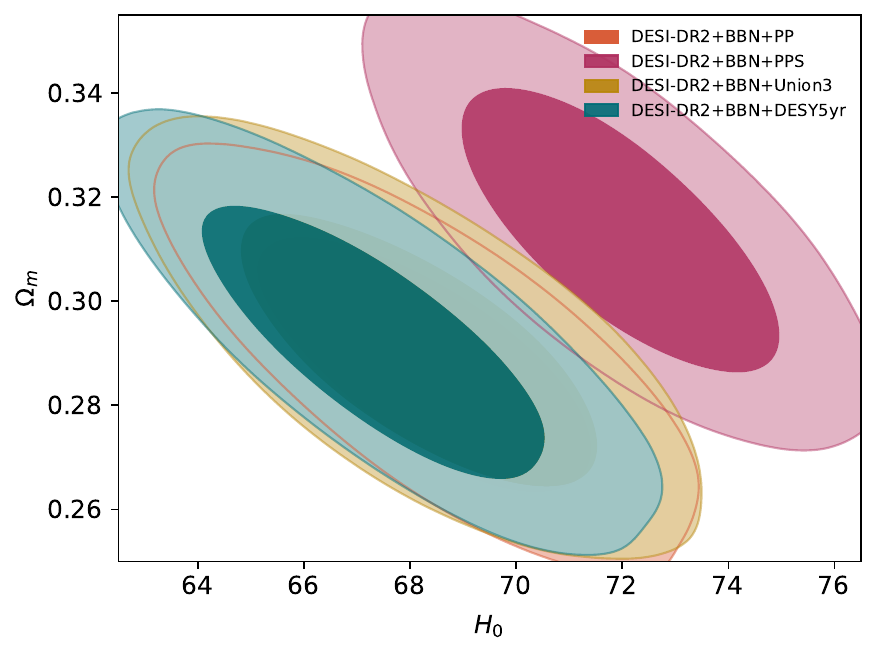}
    \caption{Two-dimensional marginalized posterior contours in the $(\Omega_m-H_0)$ plane at 68\% and 95\% CL for the Hu-Sawicki $f(R)$ gravity model using four dataset combinations: 
    DESI-DR2+BBN+PP, DESI-DR2+BBN+PPS, DESI-DR2+BBN+Union3, and DESI-DR2+BBN+DESY5.}
    \label{f6}
\end{figure}

Fig.~\ref{f6} shows the two-dimensional 68\% and 95\% confidence-level contours in the $H_0$--$\Omega_m$ plane for the Hu-Sawicki $f(R)$ model, across the four DESI-DR2+BBN+SN combinations. All four contours display a clear negative correlation between $H_0$ and 
$\Omega_m$, a generic feature of distance-based probes in which a higher matter density can be compensated by a lower expansion rate while preserving the fit to the BAO and SN distance measurements. The outlines of DESY5yr, Union3, and PP closely overlap and are focused around $H_0 \sim 66$--$70$ km/s/Mpc and $\Omega_m \sim 0.29$--$0.30$. DESY5yr occupies the cluster's leftmost, lowest-$H_0$ area. In contrast, the PPS contour is shifted to both a slightly higher $\Omega_m$ and a higher $H_0 \sim 70$--$76$ km/s/Mpc, which is consistent with the SH0ES-calibrated value reported in Table~\ref{t1}. It is distinct from the other three combinations with only a slight overlap at the 95\% level. This difference validates the fact that the $\Omega_m$--$H_0$ constraint is somewhat robust across the uncalibrated SN Ia compilations, even tho the calibrated PPS dataset consistently favors a certain region of parameter space.\\

\begin{figure}[hbt!]
    \centering
    \includegraphics[width=0.6\linewidth]{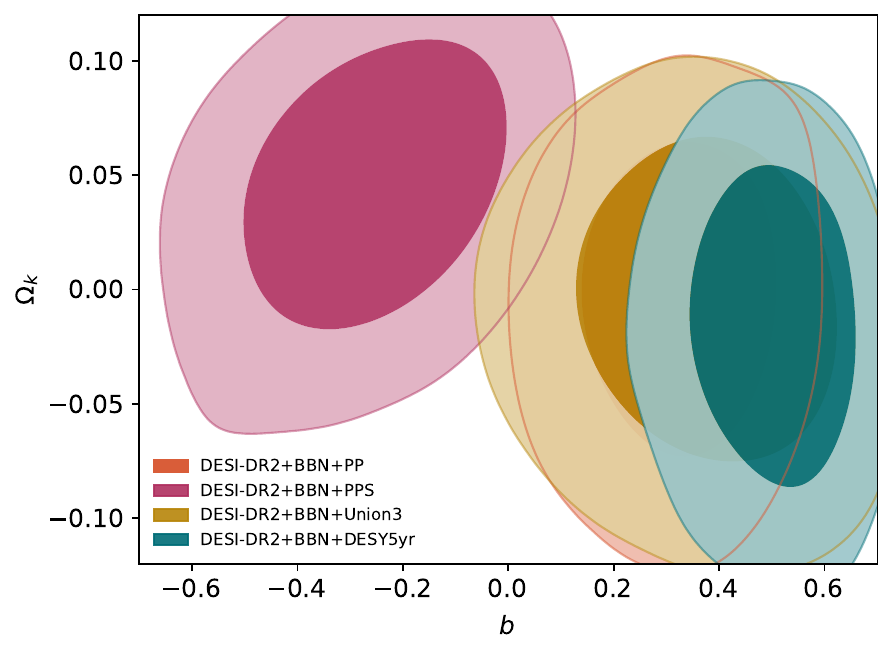}
    \caption{Two-dimensional marginalized posterior contours in the $(\Omega_k-b)$ plane at 68\% and 95\% CL for the Hu-Sawicki $f(R)$ gravity model using four dataset combinations: 
    DESI-DR2+BBN+PP, DESI-DR2+BBN+PPS, DESI-DR2+BBN+Union3, and DESI-DR2+BBN+DESY5.}
    \label{f7}
\end{figure}

Fig. \ref{f7} shows the two-dimensional 68\% and 95\% confidence-level contours in the 
$b$--$\Omega_k$ plane for the Hu-Sawicki $f(R)$ model, obtained from the four DESI-DR2+BBN+SN 
data combinations. A clear positive correlation between the modified-gravity parameter $b$ and the spatial curvature $\Omega_k$ is present in every combination, with each contour tilted 
along the same diagonal direction: larger (more positive) $b$ is accompanied by larger (more positive) $\Omega_k$, and vice versa. This behavior reflects the geometric degeneracy between the two parameters discussed in Sec.~\ref{s3}, since both act to modify the same 
distance-redshift relation probed by the BAO and SN Ia data, allowing part of the curvature signal to be reabsorbed into the modified-gravity sector.
The four contours are also well separated along the $b$-axis, reproducing the dataset 
dependence already noted in Table~\ref{t1}. The PPS combination is shifted to negative 
$b \sim -0.3$ and is the only combination whose contour lies substantially in the region 
$b<0$, consistent with the negative central value reported earlier. The Union3 and 
DESY5yr contours are instead concentrated at positive $b$, centered near $b \sim 0.3$ 
and $b \sim 0.45$--$0.5$ respectively, with DESY5yr showing the tightest and most positive 
constraint on $b$ among the four combinations. The PP contour, by contrast, appears 
only as a narrow, elongated band rather than a filled region, indicating a comparatively 
weak constraint on this parameter combination for that dataset; its 95\% contour nevertheless 
overlaps with both the Union3 and DESY5yr regions, showing that the PP data alone do not strongly discriminate between these two preferred values of $b$.
In terms of $\Omega_k$, all four contours remain broadly consistent with a flat universe 
($\Omega_k = 0$) within their 68\% regions, in agreement with the individual marginalized 
constraints listed in Table~\ref{t1}. The overlap between the PPS contour and the low-$b$ tail of 
the Union3 and DESY5yr contours, together with the shared diagonal orientation of all four ellipses, illustrates that the current combination of DESI-DR2 BAO, BBN, and SN Ia data cannot yet break the curvature-modified-gravity degeneracy in a dataset-independent way, 
reinforcing the need for tighter, calibration-independent constraints in future analyses.\\

\begin{figure}[hbt!]
    \centering
    \includegraphics[width=0.6\linewidth]{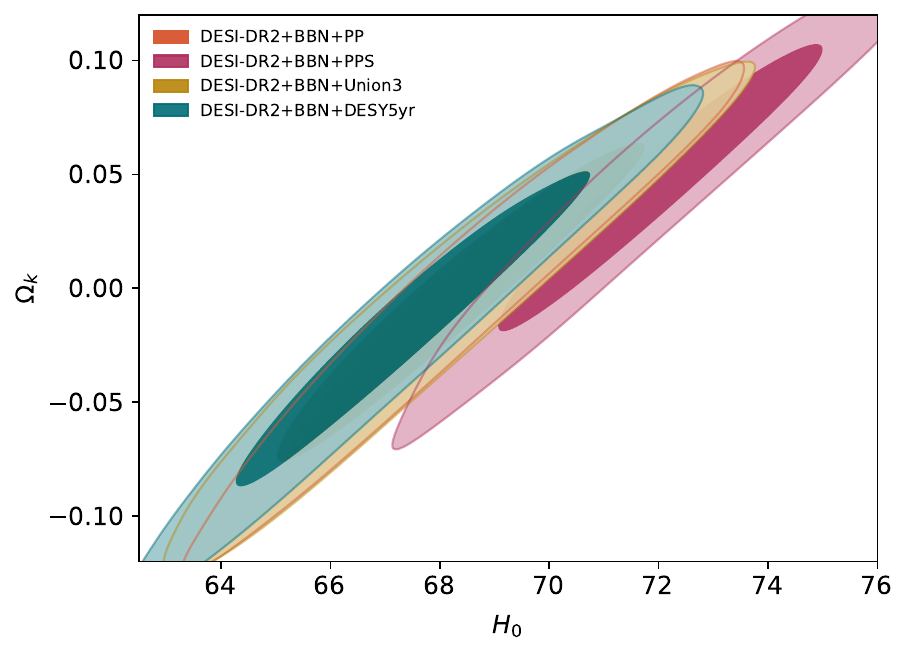}
    \caption{Two-dimensional marginalized posterior contours in the $(\Omega_k-H_0)$ plane at 68\% and 95\% CL for the Hu-Sawicki $f(R)$ gravity model using four dataset combinations: 
    DESI-DR2+BBN+PP, DESI-DR2+BBN+PPS, DESI-DR2+BBN+Union3, and DESI-DR2+BBN+DESY5.}
    \label{f8}
\end{figure}

Fig.~\ref{f8} shows the two-dimensional 68\% and 95\% confidence-level contours in the 
$H_0$--$\Omega_k$ plane for the Hu-Sawicki $f(R)$ model, for the same four DESI-DR2+BBN+SN 
combinations shown in Fig.~\ref{f7}. As with the $b$--$\Omega_k$ plane, a strong positive correlation 
between $H_0$ and $\Omega_k$ is evident in every combination, with each contour stretched 
along the same diagonal direction: higher $H_0$ is associated with higher (more positive) 
$\Omega_k$, while a lower $H_0$ pulls the geometry toward a mildly closed universe. This is the 
expected counterpart of the $b$--$\Omega_k$ degeneracy discussed for Fig.~\ref{f7}, since $b$ and 
$H_0$ are themselves correlated through the background expansion history, so any degeneracy 
involving $b$ propagates into a corresponding $H_0$--$\Omega_k$ degeneracy.
The four contours are clearly separated along the $H_0$-axis, in direct correspondence with 
the $H_0$ values reported in Table~\ref{t1}. The DESY5yr combination occupies the lowest and 
widest region, centered near $H_0 \sim 66$--$67$ km/s/Mpc, and extends furthest into negative 
$\Omega_k$, reflecting its comparatively low central $H_0$ value. The Union3 contour sits just to the right of DESY5yr, centered around $H_0 \sim 68$--$69$ km/s/Mpc, largely 
overlapping with the DESY5yr region at the 95\% level. The PPS combination  is 
shifted furthest to the right, centered near $H_0 \sim 72$--$73$ km/s/Mpc, consistent with the 
SH0ES-calibrated value in Table~\ref{t1}, and is the only combination whose contour lies mostly at 
positive $\Omega_k$. As in Fig.~7, the PP contour appears only as a thin, elongated 
band rather than a filled region, tracing the same diagonal correlation but without placing a 
strong independent constraint on either parameter individually.
Taken together, Figs.~\ref{f7} and~\ref{f8} show that the apparent detection of non-zero spatial curvature 
in some of the individual fits is not an independent result: it is tightly linked to the 
preferred values of $H_0$ and $b$ in each dataset combination, through a shared geometric 
degeneracy. This reinforces the conclusion drawn in Sec.~\ref{s3} that curvature constraints 
obtained under the Hu-Sawicki $f(R)$ model should be interpreted jointly with the modified-gravity and calibration assumptions, rather than as a standalone measurement of the 
universe's spatial geometry.\\

\textbf{Statistical Analysis: AIC and BIC:-} To assess whether the improvement in fit obtained by the Hu-Sawicki $f(R)$ model 
justifies the inclusion of the additional free parameter $b$, we perform a standard 
information-criterion comparison against $\Lambda$CDM, using the Akaike Information 
Criterion (AIC) [57] and the Bayesian Information Criterion (BIC) [58],
\begin{equation}
{\rm AIC} = \chi^2_{\rm min} + 2k, \qquad {\rm BIC} = \chi^2_{\rm min} + k \ln N,
\end{equation}
where $k$ is the number of free parameters and $N$ is the number of data points used 
in the fit. For the Hu-Sawicki model $k = 6$ ($\omega_b$, $\omega_{cdm}$, $H_0$, $b$, 
$\Omega_k$, $M_B$), while for $\Lambda$CDM $k = 5$, since the parameter $b$ is absent. 
We define $\Delta{\rm AIC} = |{\rm AIC}_{f(R)} - {\rm AIC}_{\Lambda{\rm CDM}}|$ and 
$\Delta{\rm BIC} = |{\rm BIC}_{f(R)} - {\rm BIC}_{\Lambda{\rm CDM}}|$, following the 
usual convention of reporting these differences in absolute value, and interpret them 
using the Jeffreys-type scale summarized in Ref.~[59]: $\Delta < 2$ is inconclusive, 
$2 \le \Delta < 6$ constitutes positive evidence against the disfavoured model, 
$6 \le \Delta < 10$ is strong evidence, and $\Delta \ge 10$ is decisive.

\begin{figure}[hbt!]
    \centering
    \includegraphics[width=1\linewidth]{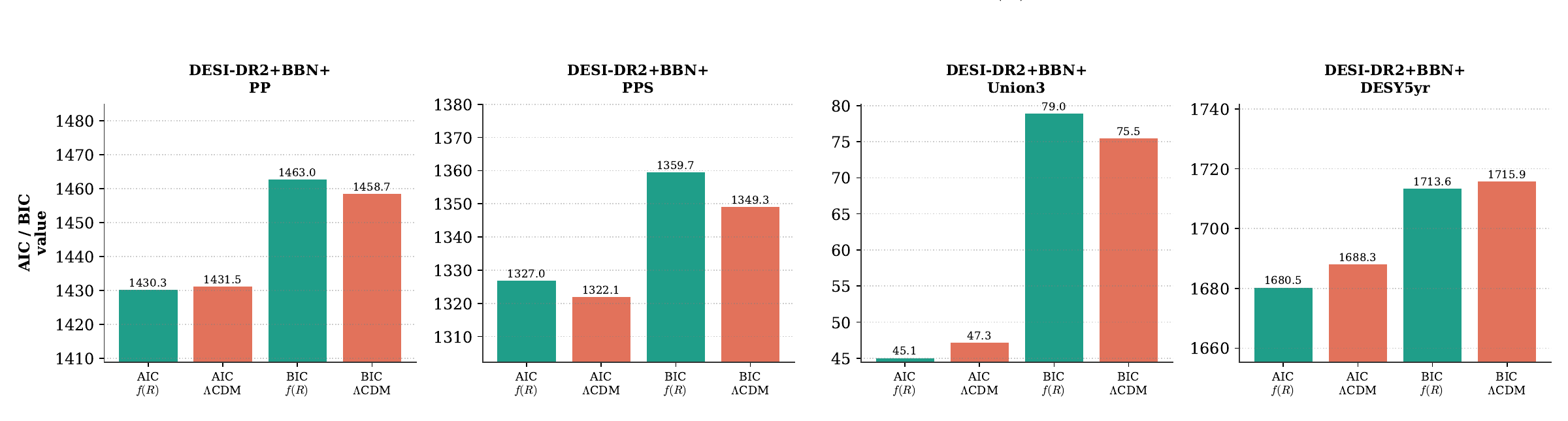}
    \caption{AIC and BIC values for the Hu-Sawicki $f(R)$ model and $\Lambda$CDM, shown separately for each of the four data combinations: DESI-DR2+BBN+PP, DESI-DR2+BBN+PPS, DESI-DR2+BBN+Union3, and DESI-DR2+BBN+DESY5yr.}
    \label{f9}
\end{figure}

\begin{figure}[hbt!]
    \centering
    \includegraphics[width=0.6\linewidth]{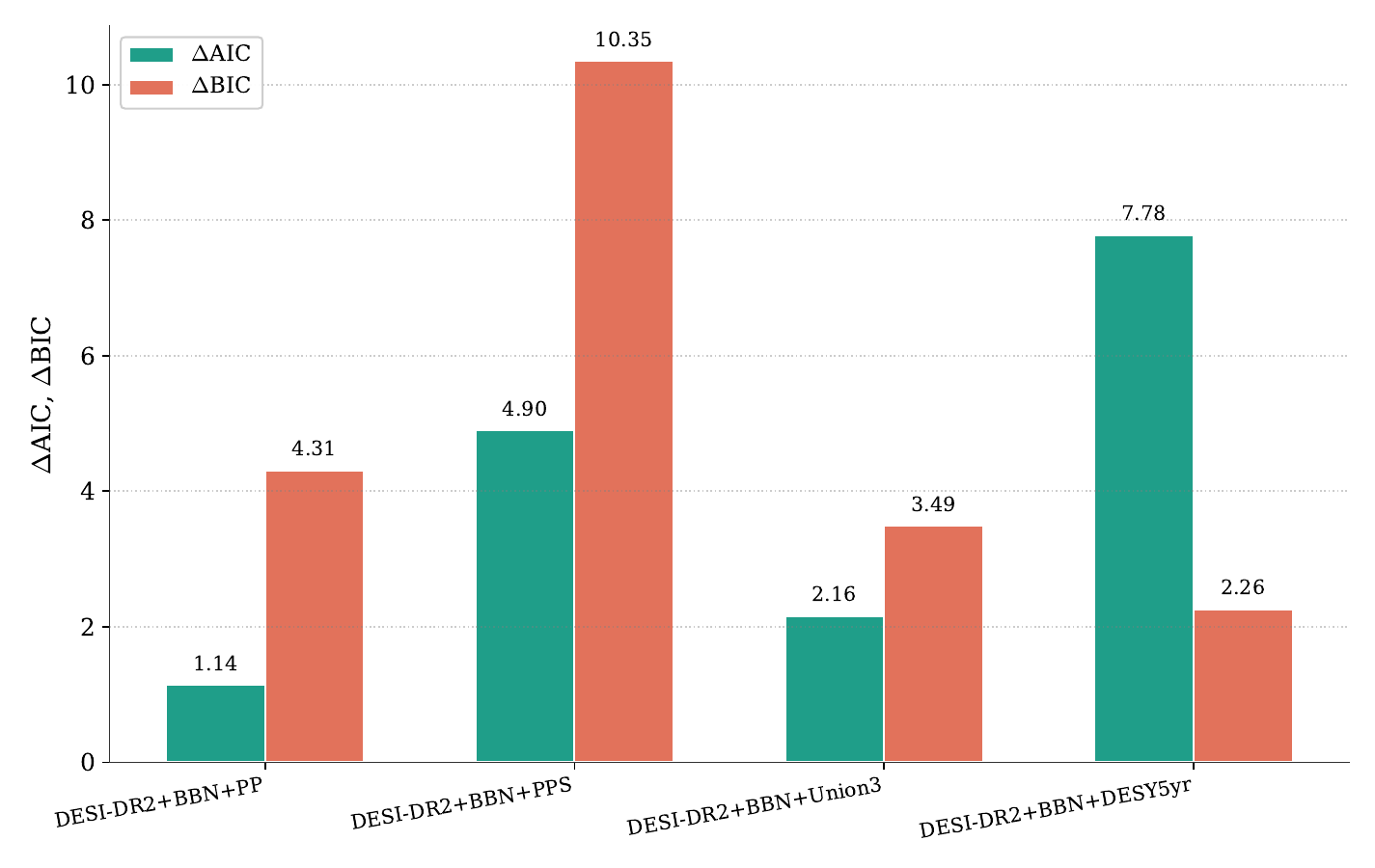}
    \caption{$\Delta$AIC and $\Delta$BIC, defined as the absolute difference between the Hu-Sawicki $f(R)$ and $\Lambda$CDM information-criterion values for each of the four data combinations. }
    \label{f10}
\end{figure}

Table~\ref{t2} lists the resulting AIC and BIC values for both models across the four data 
combinations, and Fig.~\ref{f9} displays these values graphically, together with the 
corresponding $\Delta$AIC and $\Delta$BIC in Fig.~\ref{f10}.

\begin{table}[hbt!]
\centering
\caption{AIC and BIC values for the Hu-Sawicki $f(R)$ model and $\Lambda$CDM, for each 
of the four data combinations. $\Delta$AIC and $\Delta$BIC denote the absolute 
difference between the two models.}
\begin{tabular}{l|c|c|c|c}
\hline
Data & AIC ($f(R)$/$\Lambda$CDM) & BIC ($f(R)$/$\Lambda$CDM) & $\Delta$AIC & $\Delta$BIC \\
\hline
DESI-DR2+BBN+PP        & 1430.34 / 1431.48 & 1463.03 / 1458.72 & 1.14  & 4.31  \\
DESI-DR2+BBN+PPS       & 1326.98 / 1322.08 & 1359.67 / 1349.32 & 4.90  & 10.35 \\
DESI-DR2+BBN+Union3    & 45.12 / 47.28     & 79.02 / 75.53     & 2.16  & 3.49  \\
DESI-DR2+BBN+DESY5yr   & 1680.50 / 1688.28 & 1713.62 / 1715.88 & 7.78  & 2.26  \\
\hline
\end{tabular}
\label{t2}
\end{table}

The $\Delta$AIC values indicate that the Hu-Sawicki model is essentially indistinguishable 
from $\Lambda$CDM for the PP combination ($\Delta{\rm AIC} = 1.14$), while positive evidence 
in favour of the extra parameter emerges for Union3 ($\Delta{\rm AIC} = 2.16$) and, most 
notably, for DESY5yr ($\Delta{\rm AIC} = 7.78$), which falls in the strong-evidence regime. 
The PPS combination is the only case where AIC shows positive evidence against the 
Hu-Sawicki model ($\Delta{\rm AIC} = 4.90$), consistent with its comparatively worse 
$\chi^2_{\rm min}$ noted in Sec.~\ref{s3}. The BIC values tell a different story, as expected given the stronger penalty this 
criterion imposes on additional parameters through its $\ln N$ dependence. Both PP and 
Union3 shift into the regime of positive evidence against the Hu-Sawicki model 
($\Delta{\rm BIC} = 4.31$ and $3.49$ respectively), the PPS combination moves from positive 
into strong-to-decisive evidence against it ($\Delta{\rm BIC} = 10.35$), and only the DESY5yr 
combination continues to favour the extra parameter, though the evidence weakens 
substantially compared to AIC ($\Delta{\rm BIC} = 2.26$, now only marginally above the 
inconclusive threshold). 
Consequently, whenever the improvement in $\chi^2_{\rm min}$ from including $b$ is modest, as is the case for PP and Union3, AIC still registers a mild preference for the Hu-Sawicki model while BIC does not. Only for DESY5yr, where the reduction in 
$\chi^2_{\rm min}$ relative to $\Lambda$CDM is largest ($\Delta\chi^2_{\rm min} = 9.78$), 
is the improvement strong enough to be picked up by both criteria, although even here the BIC evidence is comparatively weak. We conclude that the current DESI-DR2+BBN+SN data do 
not provide a decisive statistical preference, under either criterion, and across all four 
combinations simultaneously, for the Hu-Sawicki $f(R)$ model over $\Lambda$CDM; the 
DESY5yr combination offers the most consistent (though not decisive) support for the 
modified-gravity scenario, while the SH0ES-calibrated PPS combination consistently favors 
$\Lambda$CDM under both AIC and BIC.

\section{Conclusion}
\label{s4}

This work examined the Hu-Sawicki $f(R)$ gravity model without imposing spatial flatness, 
combining DESI-DR2 BAO, BBN, and four independent Type Ia supernova compilations -- 
PantheonPlus, PantheonPlus+SH0ES, Union3, and DESY5yr  to constrain the modified-gravity 
parameter $b$ and the curvature density $\Omega_k$ jointly rather than fixing one while 
varying the other.\\

The background parameters $H_0$ and $\Omega_m$ remain close to their $\Lambda$CDM values 
across the PP, Union3, and DESY5yr combinations, while the SH0ES-calibrated PPS dataset 
stands apart, a shift that reflects the Cepheid anchor rather than any feature specific to 
$f(R)$ gravity. The parameter $b$ shows a more interesting pattern: three of the four 
combinations favour $b \neq 0$ at better than $2\sigma$, with DESY5yr giving the clearest 
departure from $\Lambda$CDM, while PPS alone prefers a negative $b$, again traceable to its 
calibration. Since the sign and magnitude of $b$ vary with the supernova sample used, the 
present data do not yet settle the departure from $\Lambda$CDM in a dataset-independent way.\\

The curvature results are the central finding of this study. Once $b$ is allowed to vary 
freely, $\Omega_k$ is consistent with a flat universe at $1\sigma$ in every combination, 
even in the DESY5yr and PPS cases where $\Lambda$CDM fits to the same data mildly prefer an 
open or closed geometry, respectively. This reversal is explained by the strong positive 
correlations among $b$, $H_0$, and $\Omega_k$ visible in the posterior contours: curvature 
and the modified-gravity distortion parameter act on the same distance-redshift relation, so 
part of what appears as curvature under $\Lambda$CDM can instead be absorbed into $b$ once 
the extra freedom is introduced. This suggests that curvature constraints obtained under 
$\Lambda$CDM should not be interpreted independently of the assumed gravitational sector.\\

The AIC and BIC comparison reinforces the need for caution before favouring either model. 
AIC, with its lighter penalty on additional parameters, gives a mild to strong preference 
for the Hu-Sawicki model in three of the four combinations, while BIC's stronger penalty, 
driven by the large supernova sample sizes, shifts the preference back toward $\Lambda$CDM 
in most cases; only DESY5yr is favoured under both criteria simultaneously. Taken together, 
these results show that the Hu-Sawicki $f(R)$ model with free spatial curvature remains 
statistically viable and competitive with $\Lambda$CDM, without being decisively preferred 
by the current combination of DESI-DR2, BBN, and SN Ia data. Improved BAO precision from 
future DESI releases, together with tighter control over supernova calibration systematics, 
will be needed to determine whether the mild preference for $b \neq 0$ found here reflects a 
genuine departure from $\Lambda$CDM or is simply a consequence of present observational 
limitations.\\

\section*{Declaration of competing interest}
The authors declare that they have no known competing financial
interests or personal relationships that could have appeared to influence
the work reported in this paper.

\section*{Data availability}
We employed publicly available Pantheon Plus data, Pantheon Plus SH0ES 
data, BBN data, Union3, DESY5, and DESI-DR2 data are presented in 
this study. 

\section*{acknowledgments}
The author (S. Verma) is supported by a Senior Research Fellowship (UGC Ref No. 192180404148) from the University Grants Commission, Govt. of India. \\

\hspace{6cm} \textbf{Appendix I : Triangle Countor}\\

\begin{figure}[hbt!]
    \centering
    \includegraphics[width=1\linewidth]{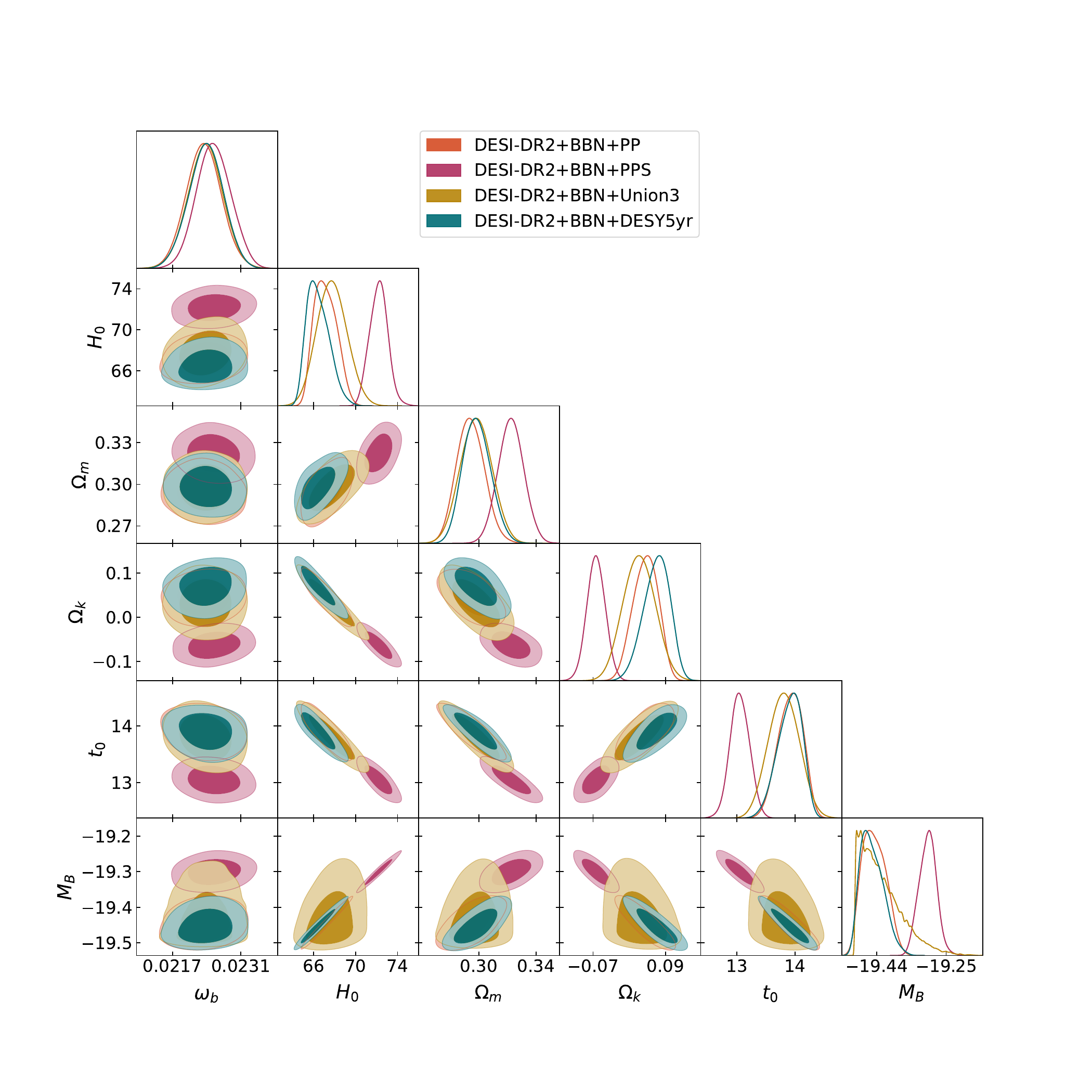}
    \caption{$\Lambda$CDM}
    \label{f11}
\end{figure}

In this appendix, we present a triangular plot with One-D posterior distributions and Two-D marginalized confidence regions ($68\%$ CL and $95\%$ CL) for all considered parameters presented in Table \ref{t1} for $\Lambda$CDM model with different combination of data sets (see Fig. \ref{f11}).

\end{document}